\documentclass[12pt]{article}

\usepackage[body={17.5cm, 23cm},right=2cm]{geometry}

\usepackage{color}
\usepackage{graphicx}
\usepackage{epsf}
\usepackage{graphicx,epsfig}
\usepackage{amsmath}
\usepackage[mathscr]{euscript}
\usepackage{pbox}
\usepackage{float}
\usepackage{amsmath}
\usepackage{amssymb}
\usepackage{epsfig}
\usepackage{cite}
\usepackage{color,colordvi}
\newcommand{\be}{\begin{eqnarray}}
\newcommand{\ee}{\end{eqnarray}}
\newcommand{\bi}{\begin{itemize}}
\newcommand{\ei}{\end{itemize}}

\newcounter{hran}

\def\MSbar{\relax\ifmmode\overline{\rm MS}\else{$\overline{\rm MS}${ }}\fi}

\begin{document}\thispagestyle{empty}

\vspace{0.5cm}

\def\thefootnote{\arabic{footnote}}
\setcounter{footnote}{0}

\def\s{\sigma}
\def\nn{\nonumber}
\def\p{\partial}
\def\ls{\left[}
\def\rs{\right]}
\def\lc{\left\{}
\def\rc{\right\}}
\def\S{\Sigma}
\def\l{\lambda}
\newcommand{\beq}{\begin{equation}}
\newcommand{\eeq}[1]{\label{#1}\end{equation}}
\newcommand{\bea}{\begin{eqnarray}}
\newcommand{\eea}[1]{\label{#1}\end{eqnarray}}

\renewcommand{\be}{\begin{eqnarray}}
\renewcommand{\ee}{\end{eqnarray}}
\renewcommand{\th}{\theta}
\newcommand{\bth}{\overline{\theta}}

\hspace*{12cm}
CERN-PH-TH-2015-130

\vspace*{2cm}

\begin{center}

{\Large \bf 
${\cal R}^2$  Supergravity
 %\\ [0.4cm]  and 
 }
\\[1.5cm]
{\normalsize   Sergio Ferrara$^{1,2,3}$,  Alex Kehagias$^{4}$ and Massimo Porrati$^{1,5}$
}
\\[1.1cm]

\vspace{.1cm}
{\small {  $^{1}$ Physics Department, Theory Unit, CERN,
CH 1211, Geneva 23, Switzerland}}\\

\vspace{.1cm}
{\small {  $^{2}$ INFN - Laboratori Nazionali di Frascati,
Via Enrico Fermi 40, I-00044 Frascati, Italy}}\\

\vspace{.1cm}
{\small {  $^{3}$Department of Physics and Astronomy, University of California
Los Angeles, CA 90095-1547, USA}}\\

\vspace{.1cm}
{\small { $^{4}$ Physics Division, National Technical University of Athens, 15780 Zografou, 
Athens, Greece}}\\

%\vspace{.1cm}
%{\small {  $^{4}$ Department of Theoretical Physics
%24 quai E. Ansermet, CH-1211 Geneva 4, Switzerland}}\\

\vspace{.1cm}
{\small {  $^{5}$
CCPP, Department of Physics, NYU 4 Washington Pl. New York NY 10003, USA}}

%\vspace{.3cm}

%\vspace{.2cm}

\end{center}

\vspace{1.2cm}

%\hrule \vspace{0.3cm}
\begin{center}
{\small  \noindent \textbf{Abstract}} \\[0.5cm]
\end{center}
\noindent 
{\small
We formulate $R^2$  pure supergravity as a scale invariant theory built only in terms of superfields 
 describing the geometry of curved superspace. 
%of the curvature chiral multiplet of ${\cal N}=1$ superspace. 
The standard supergravity
duals are obtained in both ``old" and ``new" minimal formulations of auxiliary fields. These theories have massless fields in de Sitter space as they do in their
non supersymmetric counterpart. Remarkably, the dual theory of $R^2$ supergravity in the new minimal formulation is an extension of the Freedman model,
describing a massless gauge field and a massless chiral multiplet in de Sitter space, with  inverse radius proportional to the Fayet-Iliopoulos term. This model can be interpreted as the ``de-Higgsed" phase of the
dual companion theory of $R+R^2$ supergravity.   
}
%\vspace{0.5cm}  \hrule
\vskip 2cm

\def\thefootnote{\arabic{footnote}}
\setcounter{footnote}{0}

%\vfill
%\vskip.2in
%\line(1,0){250}\\
%{\footnotesize {$^*$On leave of absence from Department of Physics and %Astronomy, University of California Los Angeles, CA 90095-1547
%USA}}

%\xfilll[-12pt]{12pt}

%\maketitle

%\date{\today}

\baselineskip= 19pt
\newpage 

\section{Introduction}

Recently, various authors \cite{KLT,KKLR,GKKLR} considered  pure  $R^2$ theories of gravity coupled to matter.  
%revived  scale invariant theories of gravity where the Einstein term is supposed to arise as a result of quantum effects.
These theories are particularly interesting also in regard to cosmology because  they naturally accommodate for de Sitter universes. While demanding conformal
invariance (Weyl local gauge symmetry) would require spin two ghosts arising from the Weyl square term \cite{stelle,GFvN}, the rigid scale invariant $R^2$ theory 
propagates only physical massless modes in
de Sitter space, in contrast with the  $R+R^2$ theory, which has in addition a Minkowski phase with a massive scalar, the inflaton. 
An Einstein term is then obtained through quantum effects as substantiated
by the analysis of \cite{KLT}.
%If an Einstein term is then obtained through quantum effects as substantiatedby the analysis of KLT.
Further restrictions that follow from the supersymmetric extensions of these theories are the aim of the present investigation.
In particular, in this note we implement the analysis of \cite{KLT} by requiring that pure $R^2$ supergravity be effectively derived solely in terms of the geometry of curved superspace.
This poses severe restrictions on the dual standard supergravity theory which, in fact, cannot be an arbitrary scale invariant theory of supergravity.
For example, we find that only certain cases are possible among the ones  worked out in \cite{KLT}.  Moreover, one  conformally coupled chiral superfield (that we call $S$) and  another one that we call the chiral superfield  $T$,  are not matter fields but have  a pure gravitational origin.
In fact, out of the three (unique) suggested forms of superpotential in
\cite{KLT}, only a certain linear combination  of $TS$ and $S^3$ can arise; $T^{3/2}$ alone is also possible. 
This result parallels  the same analysis made in the $R+R^2$ theory
\cite{Cecotti, Cecotti-Kallosh}.
Also, the $T^{3/2}$  theory, where only the $T$ field is present, is not an $R^2$ completion but a particular (scale invariant) case of the  super $F(R)=R^3$  chiral theory originally investigated in \cite{Ketov,KS1,FKP}.
The latter has in fact an anti-de Sitter  rather than a de Sitter phase 
%This theory has in fact a negative (anti de Sitter) cosmological constant rather than a positive one 
\cite{FKP,ENOl,KLT}. All the scale invariant theories discussed above present  instabilities and therefore have to be modified both at the classical and  at the quantum level.
The problem caused by these instabilities is similar to the one  found
in the context of $R+R^2$ supergravity 
\cite{ENOl,KL,EON1,FKR,FKLP,FKLP2,FKvP,FKP,KL1,EON,FeKR,ADFS,FKL1,
KL00,DZ,FP,Far,Far2,LT},
which was solved in \cite{KL}.
Even more interesting is the analysis in the new minimal formulation
\cite{Sohnius:1982fw,Ferrara:1988qxa}. Here the dual supergravity  $R+R^2$ theory is a gauge theory in the Higgs phase \cite{CFPS,FKR,FKLP,FKLP2}.
The de-Higgsed phase corresponds to the pure $R^2$ theory 
in the limit that $H M_P=g M_P^2$ is kept fixed ($H$ is the Hubble constant) while the gauge coupling goes to zero. This theory is in fact the extension of the Freedman model \cite{Freedman},
where a massless vector multiplet with a Fayet-Iliopoulos (FI) term gives rise to a positive cosmological constant. Here there is an additional massless chiral field,
dual to the antisymmetric tensor auxiliary field that has become dynamical. 

The paper is organized as follows. In section 2 we present the superconformal rules needed for our analysis and we discuss the pure $R^2$  in the old minimal formulation of the ${\cal N}=1$ supergravity. We also present the corresponding scale invariant matter couplings. In section 3, where chiral multiplets are added, we describe pure $R^2$ supergravity in the new  minimal formulation;    we conclude in section 4.

%\section{Superconformal Rules}

\section{$R^2$ Supergravity in the Old Minimal Formulation}

For our convenience we report here some rules of  superconformal tensor calculus that will be useful in order to go from the $R^2$ theory to its standard supergravity form. These rules are explained 
in \cite{FvP}, and also in \cite{FKvP,FKP}. 

Superconformal fields are denoted by their Weyl weight $w$ and chiral weight $n$. So we will use the notation $X_{w,n}$ and we will only consider scalar superfields. The basic operator is the $\Sigma$ operator, which is the curved superspace analog of $\bar D ^2$. The $\Sigma$ operator has weights $(1,3)$ and it can be applied to a superconformal field $X_{w,w-2}$ so that $\Sigma X_{w,w-2}$ is a chiral superfield of weights $(w+1,w+1)$. $X$ can be (anti)chiral only if $w=-1$, in which case $\Sigma\bar X _{(1,-1)}$ is a chiral superfield of weigh $(2,2)$. The basic identity between F and D densities of a chiral superfield $f$ of weight $(0,0)$ is 
\begin{eqnarray}
[f {\cal R} S_0^2]_F=[(f+\bar f)S_0\bar S _0]_D, \label{i1}
\end{eqnarray}
where 
${\cal R}=(\Sigma(\bar S _0)/S_0)_{(1,1)}$ is the chiral scalar curvature multiplet. 
 The  notation $[O]_{D,F}$ denotes, as usual, the
standard D- and F-term density formulae of conformal supergravity, for  a real 
superfield $O$ with scaling
weight 2 and vanishing chiral weight or a chiral superfield with Weyl (and chiral) weight 3. 
%In the old minimla formulation of ${\cal N}=1 $ supergravity, one employs a compensator chiral superfield $S_0$ as well as the
%curvature chiral superfield ${\cal R}$, both of scaling and chiral weight equal to 1. 
In particular,
the bosonic components of the  curvature 
chiral scalar multiplet  ${\cal R}$  are
\be
{\cal R}=\frac{1}{3}\bar u +\cdots+\theta^2 {\mathscr{F}}_R,
\ee
where 
\be
{\mathscr{F}}_R=-\frac{1}{2} R-3 A_\mu^2+3 i D^\mu A_\mu .
\ee
%Let us also note that ${\cal R}/S_0$ is of zero chiral and Weyl weight and its bosonic content is 
%\be
%{\cal R}/S_0=\bar X+\cdots +\theta^2 ({\mathscr{F}}_R-18 X\bar X).
%\ee
and  $u, A_\mu$ are the supergravity auxiliary fields  \cite{FvN,SW}. 
%
%\be
%X=\frac{1}{3}u=\frac{1}{3}(S-iP)\, , 
%\ee
%are the supergravity auxiliaries. 

\subsection{Scale invariant Supergavity}

It can easily be seen from the chiral curvature superfield ${\cal R}$ that we can write the following scale-invariant 
supergravity action 
\be
{\cal L}_{scal.inv}=\alpha [{\cal R}\bar {\cal R}]_D 
-\beta[{\cal R}^3]_F, \label{a1}
\ee 
where $\alpha,\beta$ are dimensionless couplings. 
We may  write eq.(\ref{a1})  in a dual form by introducing Lagrange multiplier superfields $T,S$ so that 
\be
{\cal L}_D=\alpha\big{[}S_0\bar S_0 S \bar S\big{]}_D-\beta\big{[}S_0^3S^3\big{]}_F-3\bigg[T\left(\frac{{\cal R}}{S_0}-S\right)S_0^3\bigg]_F  ,  \label{a2}
\ee
It is easy to check  that by  integrating out the Lagrange multiplier superfield $T$ in (\ref{a2}), we get back the original theory 
(\ref{a1}).
However, by using the identity in eq.(\ref{i1}),
%\cite{cecotti,FKvP}
%\be
%[T R S_0^2]_F=[(T+\bar T)S_0\bar{S}_0]_D ,
%\ee
we may write (\ref{a2}) as
\be
{\cal L}_D=-[(3T+ 3\bar T-\alpha S\bar S)S_0\bar{S}_0]_D+[ (-\beta S^3+3 T S)S_0^3]_F ,\label{a3}
\ee
which describes standard supergravity with K\"ahler potential ($\alpha\geq 0$)
\begin{eqnarray}
K=-3 \log\left(T+\bar T-\frac{\alpha}{3} S\bar S\right),
\end{eqnarray}
and superpotential
\begin{eqnarray}
W(T,S)=3 TS-\beta S^3 .
\end{eqnarray}
\vskip.1in

\noindent
{\bf The case $\mathbf{\boldsymbol\alpha=0}$}. 
In this particular case, the scale invariant supergravity action turns out to be
\be
{\cal L}_D=-\beta[{\cal R}^3]_F, \label{a30}
\ee
which can be written in a dual form as 
\be
{\cal L}_D=-[3(T+ \bar T)S_0\bar{S}_0]_D+[ (-\beta S^3+3 T S)S_0^3]_F .\label{a33}
\ee
We see that 
$S$ appears now as a Lagrange multiplier superfield and it can be   
integrated  out. As a result, we find that $S=(T/\beta)^{1/2}$ and eq.(\ref{a33}) is written as
\be
{\cal L}_D=-[3(T+ \bar T)S_0\bar{S}_0]_D+
[ 2\beta^{-1/2} T^{3/2}S_0^3]_F ,\label{a333}
\ee
so that the K\"ahler potential and the superpotential are given by 
\begin{eqnarray}
&&K=-3 \log(T+\bar T)\, , ~~~\\
&&W=\frac{2}{\beta^{1/2}}T^{3/2}.
\end{eqnarray}
This is one of the models used in \cite{KLT} to describe a supergravity dual of pure $R^2$ supergravity. However, its origin is not from the scale invariant ${\cal R}\bar {\cal R}$ term but rather from the other scale invariant ${\cal R}^3$ term. As observed in \cite{KLT}, it has a negative cosmological constant and so it cannot be the dual of an $R^2$ theory \cite{FKP,ENOl}.

\subsection{Scale Invariant Matter Couplings}

Pure $R^2$ supergravity (in its dual formulation) is invariant under scale symmetry under 
\begin{eqnarray}
T\to T e^{\lambda}, ~~~S\to S e^{\lambda/2}, ~~~S_0\to S_0 e^{-\lambda/2},
\end{eqnarray}
which is inherited from the scale symmetry of the gravitational $R^2$ theory
\begin{eqnarray}
\frac{{\cal R}}{S_0}\to \frac{{\cal R}}{S_0} e^{\lambda/2} , ~~~S_0\to S_0 e^{-\lambda/2}.  \label{rs}
\end{eqnarray} 
Let us add $n$ superconformal chiral multiplets $A^i$ with scaling 
$A^i\to A^i e^{\lambda/2}$ (but $(0,0)$ superfield weights). Then as in \cite{KLT}, we can have conformally coupled matter $C^{\bar i}_jA^j \bar A_{\bar i} S_0 \bar S _0$ but also chiral F terms coupling to the curvature  $R$~\footnote{A term $d_i R \bar A ^i$ generates a mixing $d_i S \bar A^i+h.c$ in the K\"ahler potential is also possible.} 
\begin{eqnarray}
RC_{ij} A^i A^jS_0^2,~~~\frac{R^2}{2}C_i A^i S_0,~~~C_{ijk} A^i A^j A^k S_0^3,
\end{eqnarray}
with some constant coefficient $C^{\bar i}_j,~C_{ij},~C_i, ~C_{ijk}$.
In this case,  the dual theory takes the form
\begin{eqnarray}
[(T+\bar T -\alpha S \bar S-C^{\bar j}_i A^i \bar A _{\bar j})S_0 \bar S _0]_D
+[W(T,S,A^i)S_0^3]_F
\end{eqnarray}
with \footnote{Note that the terms containing $S$ in $W$ can be transferred to the K\"ahler potential by a $T$ redefinition.}
\begin{eqnarray}
W(T,S,A^i)=-T S +\beta S^3+\frac{S^2}{2}C_i A^i+S C_{ij} A^i A^j
+C_{ijk} A^i A^j A^k.
\end{eqnarray}
This is a restricted superpotential which does not have a $T^{3/2}$ term, neither other direct coupling to matter. Note that the scaling symmetry weight
is not the same as the superconformal weight, that in our notation is always 
$(0,0)$ for all chiral fields with the exception of $S_0$ ${(1,1)}$ and ${\cal R}$ ${(1,1)}$. ${\cal R}$ is actually scale inert  as it is obvious from eq.(\ref{rs}). So the scale symmetry in the pure gravitational theory is only dictated by the compensator $S_0$.

\section{$R^2$ Supergravity in the New Minimal Formulation}

In  new-minimal supergravity, the 
appropriate gauging is implemented by a 
 real linear multiplet $L$ with scaling weight $w=2$ and vanishing chiral weight $n=0$ \cite{Sohnius:1982fw,FGKP}. In particular, the  pure $R^2$ new minimal supergravity 
 Lagrangian can be written as 
 %\cite{CFPS}
 \begin{eqnarray}
 {\cal L} = 
%-[ L V_R ]_D +
  \frac{1}{4g^2} \bigg( [ W^{\alpha}(V_R) W_{\alpha}(V_R)]_F +{\rm h.c.} \bigg), \label{NM1}
 \end{eqnarray}
where
\begin{eqnarray}
\label{VR}
V_R &=& \text{ln}\,\left(\frac{L}{S_0\overline{S}_0}\right),
\\
\label{VRW}
W_{\alpha}(V_R) &=&-\frac{1}{4} \overline{\nabla}^2 \, \nabla_{\alpha} (  V_R ) .
\end{eqnarray}
%In addition, $[\ldots]_D$ and $[\ldots]_F$ denote superconformal D- and %F-type action densities for real and chiral superfields with ($w=2, %n=0$) and ($w=3,n=2$), respectively. 
%In particular, for a vector multiplet ${\cal V}=(B,\chi,H,K,U_\mu, \lambda,D)$we have that \cite{Sohnius:1982fw,Ferrara:1988qxa}
%\begin{eqnarray}
%[V]_D=\int d^\theta E V=e(D-\frac{1}{2}\bar \psi_\mu \gamma^\mu \gamma_5 \lambda-2 (B_\mu+\frac{i}{2}\bar \psi_\mu \gamma_5 
%\chi)H^\mu\,.
%\label{v1}
%\end{eqnarray}

The Lagrangian in eq.(\ref{NM1}) is superconformally invariant and the  superconformal symmetry  can be fixed by  choosing
$L = 1. $
Then,
the superspace geometry is described by the new-minimal formulation, 
\cite{Sohnius:1982fw,Ferrara:1988qxa}, where 
%Indeed, fixing the superconformal symmetry by  $L_0=1$, we get that 
the
 graviton multiplet  $(e_\mu^a,\psi_\mu, A_\mu,B_\mu)$ consists of: the graviton $e_\mu^a$,
the gravitino $\psi_\mu$ and two auxiliary gauge fields $A_\mu$,
and $B_{\mu\nu}$ possessing  the gauge symmetry
\begin{eqnarray}
\delta A_\mu=\partial_\mu b\, , ~~~~\delta B_{\mu\nu}=\partial_\mu b_\nu-\partial_\nu b_\mu.
\end{eqnarray}
In fact, the superconformal  gauge fixing  respects 
the $U(1)_R$ $R$-symmetry of the superconformal algebra,
which is gauged by the vector 
 $A_\mu$. Then, 
$V_R$ is the gauge multiplet of the supersymmetry algebra with components (in the Wess-Zumino gauge)
\begin{eqnarray}
\label{gaugemult}
V_R=\left(A_\mu - 3H_\mu ,
- \gamma_5\gamma^\nu 
r_\nu,
-\frac{1}{2}\hat{\cal{R}} -3 H_\mu H^\mu \right), \label{vp}
\end{eqnarray}
where $r_\nu$ is the supercovariant gravitino field strength, $\hat{{\cal R}}$ is the (supercovariant) Ricci scalar
and $H^\mu$ the Hodge dual of the (supercovariant) field strength for the auxiliary two-form \cite{Ferrara:1988qxa}

\begin{eqnarray}
 H_\mu=-\frac{1}{3!} \epsilon_{\mu\nu\rho\sigma} H^{\nu\rho\sigma}\, , ~~~
 H_{\mu\nu\rho}=\partial_\mu B_{\nu\rho}+\mbox{cyclic perm. }
 \end{eqnarray} 
Obviously, $H_\mu$ is divergenceless 
\begin{eqnarray}
\nabla^\mu H_\mu=0. \label{dH}
\end{eqnarray} 
In other words, the bosonic content of the gauge multiplet $V_R$ is 
\begin{eqnarray}
V_R=(A_\mu^-,\, 0,\, -\frac{1}{2} R-3 H_\mu H^\mu), ~~~A_\mu^-=A_\mu-3 H_\mu. \label{vrb}
\end{eqnarray}

%In addition, the Lagrangian is written as 
%\begin{eqnarray}
%{\cal L}=[LV_R]_D+\frac{\alpha}{4}W^\alpha(V_R)W_\alpha(V_R)]_F+c.c\, , %~~~V_R=\ln(L/S \bar S) \label{NM1f}
%\end{eqnarray}
%where $L$ is the linear compensator of new-minimal supergravity and $S$ is chiral. 

%We may easily recognize  the first  term of  eq.(\ref{NM1}) as the Fayet-Iliopoulos term for the gauge multiplet,  whereas the second term is its standard kinetic term. These two terms will produce, among others, a $D_R+D^2_R$ factor in the bosonic action where $D_R$ is the  
% highest component  of the gauge multiplet. 
Clearly, the  F-term in eq.(\ref{NM1}) will produce the usual $D_R^2$ term in the bosonic action, where $D_R$ is the  
highest component  of the gauge multiplet. 
 Since the latter contains the scalar curvature $R$ as can be seen form eq.(\ref{vrb}), it is obvious that  eq.(\ref{NM1}) describes an $R^2$ theory \cite{CFPS,FKR,FKLP,FKLP2}.
%As a first step to write eq. (\ref{NM1}) as standard Poincar\'e %supergravity, 
%we consider $L_0$ as an unconstrained real superfield 
%(note that by employing the equation of motion for $Y_0$ we can make $L_0$ %real linear again).
%where $V_{\text{R}}$ is the curvature multiplet of new minimal supergravity %which is the gauge 
%multiplet of the R-symmetry, with field strength 
%\be
%%W_{\alpha}(V_{\text{R}}) = -\frac{i}{4} \bar \nabla^2 \nabla_\alpha %V_{\text{R}}. 
%\eea
%The  bosonic components of $V_{\text{R}} $ can be defined by projection as
%\be
%- \frac{1}{2} [\nabla_{\alpha}, \bar \nabla_{\dot \alpha} ] V_{\text{R}} | %&=& A^{-}_{\alpha \dot \alpha}
%\\
%\frac{1}{8} \nabla \bar \nabla^2 \nabla V_{\text{R}} | &=& -\frac{1}{2} (R %+6 H^2). 
%\eea
Indeed,  by employing eqs.(\ref{VRW},\ref{vp}), we find that the bosonic part of (\ref{NM1}), is written as
%\be
%e^{-1} {\cal L}_{II} = \frac{\alpha}{8} \left(R+6H^2 \right)^2  - %\frac{\alpha}{4} F^2(A^-) .
%\eea
%The starobinsky model of inflation in new-minimal supergravity then reads
\begin{eqnarray}
\label{star}
%&=&  
e^{-1}{\cal L}=
%\frac{1}{2} R+ 2A^-_\mu H^\mu + 3H_\mu H^\mu + 
\frac{1}{8g^2} \Big{(}R+6H_{\mu}H^{\mu}\Big{)}^2  - \frac{1}{4g^2} F_{\mu\nu}(A_\rho^-)F^{\mu\nu}(A_\rho^-) \,.
\end{eqnarray}
%This is  clearly an $R^2$ theory, written in a Jordan frame as the  scalar curvature is multiplied by the Jordan factor $(1+3\alpha H_\mu H^\mu)$.   
We can integrate out   $H_\mu$ after introducing  the Lagrange multipliers 
$\Lambda, a$ 
such that 
\begin{eqnarray}
\label{star1}
e^{-1} {\cal L}=  
%\frac{1}{2} R+ 2A^-_\mu H^\mu + 3H_\mu H^\mu  + 
\frac{1}{8g^2}  \Lambda \left(R+6H_\mu H^\mu\right) -  \frac{1}{32g^2}  \Lambda^2    - \frac{1}{4g^2} F_{\mu\nu}F^{\mu\nu} 
+\frac{1}{g^2}a\partial_\mu H^\mu .
\end{eqnarray}
Integrating out $\Lambda$ gives back (\ref{star}), whereas the field $a$ enforces the constraint (\ref{dH}). 
The auxiliary field $H_\mu$ appears  now quadratically  and  it can easily be integrated out   leading to
\begin{eqnarray}
H_\mu=\frac{2}{3 \Lambda}\partial_\mu a; 
\end{eqnarray}
therefore, eq.(\ref{star1}) is equivalent to
\begin{eqnarray}
\label{star2}
e^{-1} {\cal L}=  \frac{1}{8g^2}  \Lambda  R   - \frac{1}{4g^2} F_{\mu\nu}F^{\mu\nu} 
 -  \frac{1}{32g^2}  \Lambda ^2  - \frac{1}{3 g^2\Lambda}
 \partial_\mu a\partial^\mu a.
\end{eqnarray}
The theory in eq.(\ref{star2}) is  in a Jordan frame and it can be expressed in the Einstein frame after the conformal transformation 
\begin{eqnarray}
g_{\mu\nu}\to e^{-\sqrt{\frac{2}{3}}\phi} \, g_{\mu\nu}, \label{conf0}
\end{eqnarray}
where 
\begin{eqnarray}
•\phi = \sqrt{\frac{3}{2}} \, \ln \frac{ \Lambda}{4g^2}  .
\end{eqnarray}
Then, eq.(\ref{star2}),  after rescaling
  $A_\mu^- \rightarrow  g A_\mu^-$ and $a\to g^2 \sqrt{6} a$, 
  is written in the Einstein frame  as
 \begin{eqnarray}
 e^{-1} {\cal L}=  \frac{1}{2}  R   - \frac{1}{4} F_{\mu\nu}F^{\mu\nu} - 
\frac{1}{2}\p_\mu \phi \, \p^\mu \phi-\frac{1}{2}e^{-2\sqrt{\frac{2}{3}}\phi}\p_\mu a \, \p^\mu a -\frac{1}{2} g^2.  \label{fr}
 \label{suV}
 \end{eqnarray}
Therefore, $R^2$ in new minimal supergravity is described by a standard supergravity coupled to a massless vector field and a massless complex scalar 
\begin{eqnarray}
T=   \frac{1}{2}e^{\sqrt{\frac{2}{3}} \phi } + i\frac{  a}{\sqrt{6}}.
\end{eqnarray}
The field $T$ parametrizes the symmetric space $SU(1,1)/U(1)$ of scalar curvature 
 $R=-2/3$. 
%This model has a very interesting structure. 
%In superspace it is very conveniently written down in the new-minimal %framework. 
%On the other hand turning to components, 
%the non-linearity of the auxiliary sector makes 
%the theory {\it dirty} and in order to be written in a 
%simple form, 
%one has to turn to the dual standard supergravity picture. 
%It seems that a full understanding of the starobinsky model of inflation 
%requires a combined description of the minimal supergravities. 
In fact, in new minimal supergravity,  the $R^2$ theory  and its dual form, can both be described by a unique Lagrangian of the form \cite{FKLP,FKR,FP}
%Then one can check that the following Lagrangian 
\begin{eqnarray}
{\cal L}=[B(L-S_0 \bar S_0 e^U)]_D+\frac{1}{4}W^\alpha(U)W_\alpha(U)]_F+c.c.\, , \label{ld}
\end{eqnarray}
where $U$ is an unconstrained vector superfield, $W_{\alpha}(U) =-\frac{1}{4} \overline{\nabla}^2 \, \nabla_{\alpha} (  U )  $ and $B$ is a real-multiplet Lagrange multiplier. 
It is easy to see that by integrating out $B$ we find that 
\begin{eqnarray}
U=V_R.   \label{UVR}
\end{eqnarray}
 Substituting (\ref{UVR}) into (\ref{ld}), 
we get back the new minimal supergravity action (\ref{NM1}). On the other hand, integrating out  $L$ we get 
\begin{eqnarray}
B=T+\bar T,
\end{eqnarray}
where $T$ is chiral.
 Hence,   eq.(\ref{ld}) can be  written 
 in standard old minimal form as
\begin{eqnarray}
{\cal L}=-[S_0\bar S _0 e^{U}(T+\bar T)]_D+\frac{1}{4}W^\alpha(U)W_\alpha(U)]_F+c.c \, . \label{NM3}
\end{eqnarray}
We see that the K\"ahler potential is 
\begin{eqnarray}
K = -3 \ln \left[  (T+\overline{T}  )  \right], \label{kk}
\end{eqnarray}
whereas the term $e^U$ will give rise a FI term \cite{SW2,BFN}. Indeed,  in component form,   Lagrangian (\ref{NM3}) is 
\begin{align}
\label{NM6}
e^{-1}{\cal{L}}&=\frac{1}{2} R -\frac{1}{4}   
 F^{\mu\nu} F_{\mu\nu}
%-\frac{i}{4} \epsilon^{\mu\nu\rho\sigma} F_{\mu\nu} F_{\rho\sigma} 
- \frac{3}{ (T+\overline{T})^2 } 
\partial_\mu T \,\partial_\mu \bar T 
-\frac{1}{2}g^2.
\end{align}
In fact the pure $R^2$ theory  can be seen already from the $R+R^2$ theory in the new minimal supergravity, which is described by the action  \cite{FP}, by taking an appropriate limit. 
Restoring dimensions in the FI term $\xi=gM_P^2$, the limit 
is $g\to 0$ with $\xi$ fixed, which  corresponds to a de-Higgsed phase. 
The $R+R^2$ theory is described by the master action \cite{FP}
\begin{eqnarray}
 {\cal L} = 
-[S_0 \bar S_0 e^U U]_D+[B(S_0\bar S _0 e^U-L)]_D+  \frac{1}{4g^2} ( [ W^{\alpha}(V_R) W_{\alpha}(V_R)]_F +{\rm h.c.} ). \label{NM101}
 \end{eqnarray}
The rescaling 
\begin{eqnarray}
S_0\to  S_0 e^{-\lambda/2}, ~~~B\to B e^{\lambda} , ~~~L\to L e^{-\lambda}
\end{eqnarray}
clearly gives eq.(\ref{NM1}) in the $\lambda\to \infty$ limit.
 Therefore, the dual $R^2$ theory in new minimal supergravity  can be described as standard supergravity coupled to a massless chiral superfield and a  massless vector superfield with a FI term. 
This theory is an extension of the Freedman model \cite{Freedman} by a masless chiral multiplet. The latter describes a massless vector  coupled to supergravity with a positive cosmological constant.

\section{Conclusions}

Prompted by the interesting proposal of \cite{KLT,KKLR,GKKLR},  we have discussed here the supersymmetric completion of pure $R^2$ gravity. The latter is rigidly  scale invariant and propagates a massless graviton and a massless scalar on a de Sitter backgound \cite{GKKLR}, contrary to the $R+R^2$ theory which has an additional Minkowski phase with a  massive scalar (the inflaton). In ${\cal N}=1$ supergravity,  one can write two scale invariant superspace densities, an ${\cal R}\bar {\cal R}$ (D-term) and an ${\cal R}^3$ (F-term). If both terms are present, the dual theory in old-minimal formulation contains the usual scalaron field $T$ together with a conformally coupled scalar $S$ of gravitational origin. In this case the most general  superpotential turns out to be a linear combination of $ST$ and $S^3$. However, when only the ${\cal R}^3$ term is present, it turns out that $S$ is auxiliary and after integrating it out, a superpotential of the form $T^{3/2}$ arises. This theory has an anti-de Sitter rather than a de Sitter phase \cite{FKP,ENOl,KLT}. When matter fields with definite scaling are introduced, it is possible to couple them to supergravity  either by a D-term or an F-term coupling to the chiral curvature multiplet ${\cal R}$. In this case, it turns out that the  matter fields mix with $S$ but not with $T$ in the superpotential.

  A similar analysis can be done in the new minimal formulation of ${\cal N}=1$ supergravity, which reveals that the dual theory of the $R^2$ theory is described by a massless chiral multiplet together with a massless vector  multiplet with a FI term. The dual theory is thus an extension by a chiral multiplet of the Freedman model.

\vskip.5in

\noindent
{\bf {Acknowledgment}}

\vskip.1in
\noindent
We would  like to thank L. Alvarez-Gaume, F. Farakos, K. Kounnas, 
D. L\"ust, A. Riotto and A.~Van Proeyen for discussions. 
This research was implemented
under the ARISTEIA Action of the Operational Programme Education and Lifelong Learning
and is co-funded by the European Social Fund (ESF) and National Resources. 
%It is partially
%supported by European Union’s Seventh Framework Programme (FP7/2007-2013) under REA
%grant agreement n. 329083. S.F. is supported by ERC Advanced Investigator Grant n. 226455
%Supersymmetry, Quantum Gravity and Gauge Fields (Superfields). 
M.P. is supported in part by
NSF grant  PHY-1316452.

\vskip.5in
%\newpage

%%%%%%%%%%%%%%%%%%%%%%%%%%%%%%%%%%%%%%%%%%%%%%%%%%%%%%%%%%%%%%%%%%%%%%%%
\end{document}